\documentclass[11pt]{article}

\let\ssection=\section
\renewcommand{\section}{\setcounter{equation}{0}\ssection}

\newcommand{\half}{{\scriptstyle{\frac{1}{2}}}}

\newcommand{\vx}{{\bf x}}
\newcommand{\cG}{{\cal G}}
\newcommand{\br}{{\bf x}}
\newcommand{\bk}{{\bf p}}
\def\bE{{\bf E}}
\def\cA{{\cal A}}
\newcommand{\bB}{{\bf B}}

\def\p{{\partial}}
\def\vOmega{{\vec\Omega}}
\def\vp{{\bf p}}
\def\vD{{\bf D}}
\def\vj{{\bf\jmath}}
\def\vb{{\bf b}}
\def\vA{{\bf A}}
\def\vnabla{{\vec\nabla}}

\begin{document}

\setlength{\baselineskip}{16pt}

\title{Exotic galilean symmetry, non-commutativity   
\&  the Hall effect\footnote{Talk given at
the {\it  XXIII. International Conference on 
Differential Geometric Methods in Theoretical Physics.}
Aug'05, Nankai Institute of Mathematics, Tianjin, China).
}}

\author{ 
 P.~A.~Horv\'athy 
 \\ 
 Laboratoire de Math\'ematiques et de Physique Th\'eorique\\ 
 Universit\'e de Tours\\ 
 Parc de Grandmont\\ 
 F-37200 TOURS (France) 
 \\ 
 } 
\date{\today}

\maketitle

\begin{abstract}
The ``exotic'' particle model associated with the two-parameter central
extension of the planar Galilei group can be used to derive the ground
states of the Fractional Quantum Hall Effect.
Similar equations arise for a semiclassical Bloch electron.
Exotic Galilean symmetry can also be shared by Chern-Simons  field theory of the Moyal type.
\end{abstract}


\section{Introduction}
       
Recent interest in non-commuting structures stems, as it often happens, 
from far remote fields.
In high-energy physics, it comes from the theory of
strings and membranes \cite{string}, or 
from studying  galilean symmetry in the plane
\cite{Grigore,exotic,DH,HMSvort}. 
Independently and around the same time, very similar
structures were considered in condensed matter physics, namely for the semiclassical dynamics of a Bloch electron \cite{Niu}.
Recent developments include the Anomalous \cite{AHE},
the Spin \cite{SpinHall}
and the Optical  \cite{Optical} Hall effects.

Below we first review the exotic point-particle model of \cite{DH},
followed by a brief outline of the semiclassical Bloch electron
\cite{Niu}.

Our present understanding of the Fractional
Quantum Hall Effect is
based on the motion of charged vortices in a magnetic field \cite{QHE,SCZ}. Such vortices arise as exact solutions
in a field  theory of matter coupled to an abelian gauge field $A_\nu$,
whose dynamics is governed by the Chern-Simons term \cite{CSvortex,JP}.
Such theories can be either relativistic or nonrelativistic.
In the latter case, boosts commute.
 Exotic Galilean symmetry
can be found, however, in a Moyal-version of Chern-Simons
field-theory \cite{HMSvort},  presented  in Section \ref{NCCS}.

\section{``Exotic'' mechanics in the plane}

It has been known for (at least) 33 years that the planar Galilei 
group
admits an ``exotic'' two-parameter central extension \cite{exotic}: 
unlike  in $D\geq3$ spatial dimensions,
the commutator of galilean boosts yields a new central charge,
\begin{equation}
[\cG_1,\cG_2]=\kappa.
\label{exorel}
\end{equation}

This has long remained a sort of mathematical curiosity, though. 
It has been around 1995 that people started to inquire about the physical consequences of such an extended symmetry.
In \cite{Grigore,DH}, in particular, Souriau's ``orbit method'' \cite{SSD}
was used to construct a classical system with such an
exotic symmetry. The latter is realized by the usual
galilean generators, except for the boost and the 
angular momentum,
\begin{equation}
    \begin{array}{ll}
    &j=\epsilon_{ij}x_ip_j+\half\theta\,p_ip^i,
    \\[6pt]
    &\cG_{i}=mx_{i}-p_{i}t+m\theta\,\epsilon_{ij}p_{j}.
    \end{array}
\label{consquant}
\end{equation}

The resulting free model moves, however, exactly as
in the standard case. The ``exotic'' structure behaves hence
roughly as spin:
it contributes to some conserved quantities, but the
new terms are separately conserved.
The new structure does not 
seem to lead to any new physics.

The situation changes dramatically if the particle 
is coupled to a gauge field. The resulting equations of motion read
\begin{equation}
\begin{array}{rcl}\displaystyle
m^*\dot{x}_{i}
&=&
p_{i}-\displaystyle em\theta\,\varepsilon_{ij}E_{j},
\\[8pt]
\displaystyle
\dot{p}_{i}
&=&
eE_{i}+eB\,\varepsilon_{ij}\dot{x}_{j},
\end{array}
\label{DHeqmot}
\end{equation}
where $\theta=k/m^2$ is the non-commutative parameter and
we have introduced the \textit{effective mass}
\begin{equation}
m^*=m(1-e\theta B).
\end{equation}
\goodbreak 

The changes, crucial for physical applications, are two-fold:
Firstly, the relation between velocity and momentum, (\ref{velrel}),
contains an  ``anomalous'' term so that  
$\dot{x}_i$ and $p_{i}$ are not parallel. 
The second novelty is the interplay between the exotic structure and the magnetic field, yielding the effective mass $m^*$ in (\ref{Lorentz}).

The equations (\ref{DHeqmot}) come from the Lagrangian
\begin{equation}
 \int\!(\vp-\vA\,)\cdot d\vx
-\frac{p^2}{2}\,dt
+
\frac{\theta}{2}\,\vp\times d\vp.
\label{exolag}
\end{equation}

When $m^*\neq0$,
\ref{DHeqmot} is also a Hamiltonian system, $\dot{\xi}=\{h,\xi^\alpha\}$,
with $\xi=(p_i,x^j)$ and Poisson brackets
\begin{equation}
\begin{array}{lll}
\{x_{1},x_{2}\}=
\displaystyle\frac{m}{m^*}\,\theta,
	\\[3mm]
	\{x_{i},p_{j}\}=\displaystyle\frac{m}{m^*}\,\delta_{ij},
	\\[3mm]
	\{p_{1},p_{2}\}=\displaystyle\frac{m}{m^*}\,eB,
\end{array}
\label{exocommrel}
\end{equation}

A most remarkable property is that for vanishing effective mass
$m^*=0$ i.e. when the magnetic field takes the
critical value
\begin{equation}
B=\frac{1}{e\theta},
\end{equation}
then the system becomes singular. Then ``Faddeev-Jackiw'' (alias symplectic)
reduction yields an  essentially two-dimensional, simple system,
similar to 
``Chern-Simons mechanics'' \cite{DJT}. The symplectic plane plays, simultaneously,
the role of both configuration and phase space. The only motions are those which follow a generalized Hall law,
and the quantization of the reduced system yields the ``Laughlin
wave functions'' \cite{QHE}, which are in fact the ground states
in the Fractional Quantum Hall Effect (FQHE).

The relations (\ref{exocommrel}) diverge as $m^*\to0$,
but after reduction we have, cf. (\ref{exorel}),
\begin{equation}
\{x_1,x_2\}=\theta.
\label{NCrel}
\end{equation}

\section{Semiclassical Bloch electron}

Quite remarkably, around the same time and with no relation to the
above developments, 
a very similar theory has arisen in solid state physics \cite{Niu}. 
Applying a Berry-phase argument to a Bloch electron in a lattice,
 a semiclassical model can be derived.
The equations of motion in the $n{}^{th}$ band read 
\begin{eqnarray}
\dot{\br}&=\displaystyle\frac{\p\epsilon_n(\bk)}{\p\bk}-\dot{\bk}\times\vOmega(\bk),\label{velrel}
\\[6pt]
\dot{\bk}&=-e\bE-e\dot{\br}\times\bB(\br),
\label{Lorentz}
\end{eqnarray}
where $\br=(x^i)$ and $\bk=(p_j)$ denote the electron's three-dimensional
intracell position and quasimomentum,
respectively, $\epsilon_n(\bk)$ is the band energy. The purely momentum-dependent 
$\vOmega$ is the Berry curvature of the electronic Bloch states,
$\Omega_i(\bk)=\epsilon_{ijl}\p_{\bk_j}\cA_l(\bk)$, where 
$\cA$ is the Berry connection. 

Recent applications of the model, based on the anomalous velocity term
in (\ref{velrel}), include the Anomalous
\cite{AHE} and the Spin \cite{SpinHall} 
 Hall Effects.
  
Eqns. (\ref{velrel}-\ref{Lorentz})  derive from the Lagrangian
\begin{equation}
     L^{Bloch}=\big(p_{i}-eA_{i}(\br,t)\big)\dot{x}^{i}-
     \big(\epsilon_n(\bk)-eV(\br,t)\big)
     +\cA^{i}(\bk)\dot{p}_{i},
     \label{blochlag}
\end{equation}
and are also consistent with the Hamiltonian structure
\cite{BeMo,DHHMS}
\begin{eqnarray}
\{x^i,x^j\}^{Bloch}&=\displaystyle\frac{\varepsilon^{ijk}\Omega_k}{1+e\bB\cdot\vOmega},
\label{xx}
\\[6pt]
\{x^i,p_j\}^{Bloch}&=\displaystyle\frac{\delta^{i}_{\ j}
+eB^i\Omega_j}{1+e\bB\cdot\vOmega},
\label{kk}
\\[6pt]
\{p_i,p_j\}^{Bloch}&=-\displaystyle\frac{\varepsilon_{ijk}eB^k}{1+e\bB\cdot\vOmega}
\label{xk}
\end{eqnarray}
and Hamiltonian $h=\epsilon_n-eV$.

Restricted to the plane, these equations reduce, furthermore,
to the exotic equations (\ref{DHeqmot}) provided 
$\Omega_i=\theta\delta_{i3}$. For $\epsilon_n(\bk)=\bk^2/2m$
and
chosing $\cA_i=-({\theta}/{2})\epsilon_{ij}p_j$, 
the semiclassical Bloch Lagrangian
(\ref{blochlag}) becomes the ``exotic'' expression (\ref{exolag}).
The
exotic galilean symmetry is lost, however, if $\theta$ is not constant.

\section{Non-commutative Chern-Simons theory}\label{NCCS}

Field  theory coupled to an abelian gauge field $A_\nu$,
whose dynamics is governed by the Chern-Simons term 
admits exact  vortex solutions \cite{CSvortex,JP}.
Such theories can be either relativistic or nonrelativistic. In 
the latter case \cite{JP}, 
\begin{equation}
 L=L_{matter}+L_{field}=
    i\bar{\psi} D_{t}\psi-\half\big|{\vD\psi}\big|^2
    +\mu\left(\half\epsilon_{ij}
    \p_{t}A_{i}A_{j}+A_{t}B\right),
    \label{CSlag}
 \end{equation}
[plus a potential $U(\psi)$], where $D_\nu=\p_\nu-ieA_\nu$, $\nu=t,i$. 
Infinitesimal galilean boosts, implemented conventionally as
\begin{eqnarray}
    \delta^{0}_{}\psi&=&i\vb\cdot\vx\,\psi-t\vb\cdot\vnabla\psi,
    \label{psiCimp}
    \\[2pt]
    \delta^{0}_{}A_{i}&=&-t\vb\cdot\vnabla A_{i},
    \label{ACimp}
    \\[2pt]
    \delta^{0}_{}A_{t}&=&-\vb\cdot\vA-t\vb\cdot\vnabla A_{t},
    \label{A0Cimp}
\end{eqnarray}
are generated by the constants of the motion
\begin{eqnarray}
       \cG^0_i=t{\cal P}_{i}-\int\!x_i\,\vert\psi\vert^2\, d^2\vx,
\qquad
        {\cal P}_{i}=\int\!\frac{1}{2i}
    \big(\bar{\psi}\p_{i}\psi-(\overline{\p_{i}\psi})\psi\big) d^2\vx
    -\frac{\mu}{2}\int\!\epsilon_{jk}A_{k}\p_{i}A_{j} d^2\vx.      
\label{cboost}
\end{eqnarray}

The galilean symmetry extends in fact into a
Schr\"odinger symmetry \cite{JP}; 
there is no sign of ``exotic'' galilean symmetry, however, since
$\{\cG_1^0,\cG_2^0\}=0$. 
Replacing ordinary products
with the Moyal star-product, 
\begin{equation}
\big(f\star g\big)(x_1, x_2)=\exp\left(i\frac{\theta}{2}\big(
\p_{x_1}\p_{y_2}-\p_{x_2}\p_{y_1}\big)\right)
f(x_1, x_2)g(y_1, y_2)\Big|_{\vx=\vec{y}}
\label{thetaMoyal}
\end{equation}
where  $\theta$ is a real parameter,
a non-commutative version of the theory can
be constructed, though. The classical Lagrangian is formally
still (\ref{CSlag}), but the covariant derivative, the field strength, and the Chern-Simons term,
\begin{eqnarray}
    D_\mu\psi&=&\p_\mu-ieA_\mu\star\psi,
    \\[8pt]
    F_{\mu\nu}&=&\p_\mu A_\nu-\p_\nu A_\mu-ie
    \big(A_\mu\star A_\nu-A_\nu\star A_\mu\big),
\\[8pt]
\hbox{Chern-Simons term}&=&\frac{\mu}{2}\,\epsilon_{\mu\nu\sigma}
\left(A_\mu\star\p_\nu A_\sigma-\frac{2ie}{3}
A_\mu\star A_\nu\star A_\sigma\right),
\end{eqnarray}
respectively, all involve the Moyal form.
The variational equations read
\begin{eqnarray}
    iD_{t}\psi+\frac{1}{2}{\vD}^2\psi&=&0,\label{NCNLS}
    \\[2pt]    
   \kappa E_{i}-{e}\epsilon_{ik}j^{l}_{\ k}&=&0,\label{NCFCI}
    \\[2pt]
    \kappa B+e\rho^{l}&=&0,\label{NCGauss}
\end{eqnarray}
where $B=\epsilon_{ij}F_{ij}$,  $E_{i}=F_{i0}$, and
$\rho^l$ and $\vj{\ }^l$ denote the {\it left density} 
and {\it left current}, respectively,
\begin{eqnarray}
    \rho^{l}=\psi\star\bar\psi,
    \qquad
    {\vj}\strut{\,}^l=\frac{1}{2i}\left(\vD\psi\star\bar\psi
    -\psi\star(\overline{\vD\psi})\right).
    \label{ldenscur}
\end{eqnarray}

These equations admit, just like their ordinary counterparts,
exact vortex solutions \cite{LMSCS}.

 The modified theory is {\it not} invariant w. r. t.  boosts
implemented as above.
Galilean invariance can be restored, however, by implementing 
boosts rather as 
\begin{equation}
    \delta\psi=\psi\star(i\vb\cdot\vx)-t\vb\cdot\vnabla\psi
    =
    (i\vb\cdot\vx)\psi+\frac{\theta}{2}\vb\times\vnabla\psi
    -t\vb\cdot\vnabla\psi,
    \label{NCimp}
\end{equation}
supplemented by (\ref{ACimp})-(\ref{A0Cimp}). 
Then the generators,
\begin{equation}
    \cG_i=t{\cal P}_i-\int\!x_i\bar{\psi}\star\psi\, d^2\vx,
    \label{rncboost}
\end{equation}
do satisfy the ``exotic'' relation (\ref{exorel}) 
\begin{equation}
[\cG_1,\cG_2]=\kappa
\qquad\hbox{with}\qquad
\kappa=-\theta\int\vert\psi\vert\, d^2\vx.
\end{equation}


\noindent{\bf Acknowledgments}.
 This review is based on joint research with
C. Duval, Z. Horv\'ath, L. Martina, M. Plyushchay and
P. Stichel to whom I express my indebtedness.
I would like to thank Prof. Mo-lin Ge for his  warm
hospitality at the Nankai Institute of Mathematics
at Tianjin (China).

\goodbreak

\end{document}